\documentclass[%
 reprint,
 amsmath,amssymb,
 aps,
 prb,
 floatfix,
]{revtex4-2}

\usepackage{graphicx}
\usepackage{dcolumn}
\usepackage{bm}

\begin{document}

\title{
Composite spin and quadrupole wave in the ordered phase of Tb$_{2+x}$Ti$_{2-x}$O$_{7+y}$
}

\author{H. Kadowaki}
\affiliation{Department of Physics, Tokyo Metropolitan University, Hachioji-shi, Tokyo 192-0397, Japan}

\author{H. Takatsu}
\affiliation{Department of Physics, Tokyo Metropolitan University, Hachioji-shi, Tokyo 192-0397, Japan}

\author{T. Taniguchi}
\affiliation{Department of Physics, Tokyo Metropolitan University, Hachioji-shi, Tokyo 192-0397, Japan}

\author{B. F{\aa}k}
\affiliation{Institute Laue Langevin, BP156, F-38042 Grenoble, France}

\author{J. Ollivier}
\affiliation{Institute Laue Langevin, BP156, F-38042 Grenoble, France}

\date{\today}

\begin{abstract}
The hidden ordered state of the frustrated pyrochlore oxide 
Tb$_{2+x}$Ti$_{2-x}$O$_{7+y}$ is possibly one of the two electric 
multipolar, or quadrupolar, states of the effective pseudospin-1/2 
Hamiltonian derived from crystal-field ground state doublets of 
non-Kramers Tb$^{3+}$ ions. 
These long-range orders are antiparallel or parallel alignments 
of transverse pseudospin components 
representing electric quadrupole moments, 
which cannot be observed as magnetic Bragg reflections 
by neutron scattering. 
However pseudospin waves of these states are composite 
waves of the magnetic-dipole and electric-quadrupole moments, 
and can be partly observed by inelastic magnetic neutron scattering. 
We calculate these spin-quadrupole waves using linear 
spin-wave theory and discuss previously observed low-energy magnetic 
excitation spectra of a polycrystalline sample 
with $x = 0.005$ ($T_{\text{c}} = 0.5 $ K).
\end{abstract}

\maketitle

\section{Introduction}
Geometrically frustrated magnets have been actively studied 
in recent years \cite{Lacroix11}. 
In particular, pyrochlore magnets \cite{Gardner10} showing 
spin ice behavior \cite{Bramwell01} have interesting features 
such as finite zero-point entropy and emergent magnetic 
monopole excitations \cite{Castelnovo08}. 
A quantum spin-liquid state is 
theoretically predicted for certain spin-ice like 
systems \cite{Hermele04,Savary12,Lee12,Kato14,Gingras14}, 
where transverse spin interactions transform 
the classical spin ice into quantum spin liquid. 
This quantum spin ice (QSI), or U(1) quantum spin liquid, 
is characterized by an emergent U(1) gauge field fluctuating down to $T=0$ 
and by excitations of gapped bosonic spinons and gapless photons \cite{Hermele04,Lee12,Benton12}. 
By changing the interactions of the QSI in some ways the system undergoes a 
quantum phase transition to long range ordered (LRO) states of 
transverse spin or pseudospin \cite{Savary12}, 
being interpreted as Higgs phases \cite{Lee12,Chang12}.
Experimental investigations of the U(1) quantum spin liquid 
and neighboring LRO states have been challenged by 
several groups \cite{Chang12,Ross11,Gingras14}. 
However it is difficult to characterize the quantum spin liquid states, 
which preclude standard techniques of observing 
magnetic Bragg reflections and magnons. 

Among magnetic pyrochlore oxides \cite{Gardner10}, 
R$_2$Ti$_2$O$_7$ (R = Dy, Ho) are the well-known 
classical Ising spin-ice examples \cite{Bramwell01}. 
A similar system Tb$_{2}$Ti$_{2}$O$_{7}$ (TTO) has 
attracted much attention, 
because magnetic moments remain dynamic 
with short range correlations down to 50 mK \cite{Gardner99}. 
Since TTO has been thought to be close to the classical spin ice, 
the low-temperature dynamical behavior of TTO could be attributed to QSI \cite{Molavian07}. 
Inspired by this intriguing idea, 
many experimental studies of TTO have been performed 
to date \cite{Takatsu12,Taniguchi13,Petit12,Fennell14,Fritsch13,Fritsch14} 
(and references in Refs.~\cite{Gingras14,Petit15}). 
However the interpretation of experimental data has been a conundrum \cite{Gingras14,Petit15}, 
partly owing to strong sample dependence \cite{Chapuis09URL,Takatsu12,Taniguchi13}. 
Among these studies, our investigation \cite{Taniguchi13} 
of polycrystalline Tb$_{2+x}$Ti$_{2-x}$O$_{7+y}$ 
showed that a very small change of $x$ induces a quantum phase 
transition between a spin-liquid state ($x < -0.0025 = x_{\text{c}} $) 
and a LRO state with a hidden order parameter ($x_{\text{c}} < x$). 
It is important to clarify the origin of this order parameter, 
which becomes dynamical in the spin-liquid state ($x < x_{\text{c}} $). 

In this and companion \cite{Takatsu15,Wakita15} work, 
we try to reformulate the problem of TTO 
and to reinterpret its puzzling experimental data 
based on the theoretically predicted \cite{Onoda11} 
electronic superexchange interactions. 
A novel ingredient of these interactions is the Onoda-type 
coupling \cite{Onoda11} between neighboring electric quadrupole moments 
of non-Kramers Tb$^{3+}$ ions. 
The theory \cite{Onoda11} proposes an effective pseudospin-1/2 
Hamiltonian described by the Pauli matrices representing 
both magnetic-dipole and electric-quadrupole moments. 
Depending on the parameters of the Hamiltonian there are 
two electric quadrupole ordering phases, which are candidates for 
the hidden order of TTO. 
These electric quadrupolar orders do not bring about observable 
magnetic Bragg peaks. 
However, these orders can be detected by their elementary 
excitations (inelastic magnetic scattering), 
and by proper interpretation using a linear spin-wave 
theory. 

In this paper, starting from the crystal-field (CF) ground 
state doublet of TTO, we account for its single-site 
electric quadrupole moments, their LRO, and pseudospin 
wave excitations in the electric quadrupole LRO. 
A standard linear spin-wave theory 
predicts that the pseudospin wave in the electric quadrupole LRO 
is, in reality, a composite wave of magnetic-dipole and electric-quadrupole moments.
We discuss this possibility for Tb$_{2+x}$Ti$_{2-x}$O$_{7+y}$ 
using previously observed \cite{Taniguchi13} low-energy magnetic 
excitation spectra of a polycrystalline sample 
with $x = 0.005$ ($T_{\text{c}} = 0.5 $ K). 

\section{Crystal Field and Electric Multipole Moment}
The CF states and inelastic neutron excitation spectra of 
TTO have been investigated by many 
authors \cite{Gingras00,Mirebeau07,Bertin12,Zhang14,Princep15}; 
readers are referred to Ref.~\cite{Princep15} 
for details. 
In a low energy range, there are four CF states: 
ground doublet states and first-excited doublet states 
at $E \sim 16$ K. 
Since the interesting temperature range is below 1 K, 
we neglect the first-excited doublet states and 
consider only the ground state doublet, for simplicity. 

Among studies of CF, we adopt the CF parameters of 
Ref.~\cite{Mirebeau07} (or Ref.~\cite{Bertin12}). 
The CF ground state doublet of TTO can be written by 
\begin{equation}
| \pm 1 \rangle_{\text{D}} 
= A | \mp 4 \rangle \pm B | \mp 1 \rangle + C | \pm 2 \rangle \mp D | \pm 5 \rangle ,
\label{GD}
\end{equation}
where $| m \rangle$ stands for the $| J=6, m \rangle$ state within 
a $JLS$-multiplet \cite{Jensen91}. 
The coefficients \cite{Mirebeau07} of Eq.~(\ref{GD}) are 
$A=0.9581$, $B=0.1284$, $C=0.1210$, $D=0.2256$. 
The local symmetry axes \cite{Bertin12,Onoda11} 
of the crystallographic four sites are 
\begin{equation}
\bm{x}_0 = \tfrac{1}{\sqrt{6}}(1,1,\bar{2}), 
\bm{y}_0 = \tfrac{1}{\sqrt{2}}(\bar{1},1,0), 
\bm{z}_0 = \tfrac{1}{\sqrt{3}}(1,1,1)
\label{Laxis0}
\end{equation}
for sites at $\bm{t}_n + \bm{d}_0$ with $\bm{d}_0=\tfrac{1}{4}(0,0,0)$,
\begin{equation}
\bm{x}_1 = \tfrac{1}{\sqrt{6}}(1,\bar{1},2), 
\bm{y}_1 = \tfrac{1}{\sqrt{2}}(\bar{1},\bar{1},0), 
\bm{z}_1 = \tfrac{1}{\sqrt{3}}(1,\bar{1},\bar{1})
\label{Laxis1}
\end{equation}
for sites at $\bm{t}_n + \bm{d}_1$ with $\bm{d}_1=\tfrac{1}{4}(0,1,1)$,
\begin{equation}
\bm{x}_2 = \tfrac{1}{\sqrt{6}}(\bar{1},1,2), 
\bm{y}_2 = \tfrac{1}{\sqrt{2}}(1,1,0), 
\bm{z}_2 = \tfrac{1}{\sqrt{3}}(\bar{1},1,\bar{1})
\label{Laxis2}
\end{equation}
for sites at $\bm{t}_n + \bm{d}_2$ with $\bm{d}_2=\tfrac{1}{4}(1,0,1)$,
\begin{equation}
\bm{x}_3 = \tfrac{1}{\sqrt{6}}(\bar{1},\bar{1},\bar{2}), 
\bm{y}_3 = \tfrac{1}{\sqrt{2}}(1,\bar{1},0), 
\bm{z}_3 = \tfrac{1}{\sqrt{3}}(\bar{1},\bar{1},1)
\label{Laxis3}
\end{equation}
for sites at $\bm{t}_n + \bm{d}_3$ with $\bm{d}_3=\tfrac{1}{4}(1,1,0)$, 
where $\bm{t}_n$ is an FCC translation vector.

In the CF ground state doublet of Eq.~(\ref{GD}), the magnetic-dipole 
and electric-multipole moment operators \cite{Kusunose08} 
are represented by $2 \times 2$ matrices: 
the Pauli matrices $\sigma^x$, $\sigma^y$, $\sigma^z$ and the unit matrix. 
The magnetic dipole moment operators within $| \pm 1 \rangle_{\text{D}} $ are 
\begin{eqnarray}
J_x &=& J_y = 0,\nonumber\\
J_z &=& - (4A^2+B^2-2C^2-5D^2) \sigma^z = - 3.40 \sigma^z, 
\label{mag}
\end{eqnarray}
which implies that Tb$^{3+}$ magnetic dipole moments behave 
as Ising-like spins. 

As pointed out in Ref.~\cite{Onoda11}, 
for non-Kramers ions in the pyrochlore structure 
including Tb$^{3+}$ in TTO 
the CF ground doublet states have additionally 
electric multipole moments. 
These electric multipole moment operators are represented by 
$\sigma^x$, $\sigma^y$, and the unit matrix. 
Using the explicit form of Eq.~(\ref{GD}), 
the electric quadrupole moment operators \cite{Kusunose08} 
within $| \pm 1 \rangle_{\text{D}} $ 
are expressed by 
\begin{eqnarray}
\tfrac{1}{2}[ 3 J_z^2 -J(J+1)] &=& 
3A^2 - \tfrac{39}{2} B^2 - 15 C^2 + \tfrac{33}{2} D^2 \nonumber\\
&=& 3.05,\nonumber\\
%
\tfrac{\sqrt{3}}{2}[ J_x^2 - J_y^2] &=& 
\left( -\tfrac{21 \sqrt{3}}{2} B^2 + 9 \sqrt{10} AC \right) \sigma^x \nonumber\\
&=& 3.00 \sigma^x ,\nonumber\\
\tfrac{\sqrt{3}}{2}[ J_x J_y + J_y J_x ] &=& 
- \left( -\tfrac{21 \sqrt{3}}{2} B^2 + 9 \sqrt{10} AC \right) \sigma^y \nonumber\\
&=& - 3.00 \sigma^y ,\nonumber\\
\tfrac{\sqrt{3}}{2}[ J_z J_x + J_x J_z ] &=& 
- \left( 3 \sqrt{30} BC + 9 \sqrt{\tfrac{33}{2}} AD \right) \sigma^x \nonumber\\
&=& -8.16 \sigma^x ,\nonumber\\
\tfrac{\sqrt{3}}{2}[ J_y J_z + J_z J_y ] &=& 
- \left( 3 \sqrt{30} BC + 9 \sqrt{\tfrac{33}{2}} AD \right) \sigma^y \nonumber\\
&=& - 8.16 \sigma^y .
\label{quad2}
\end{eqnarray}
Similarly we can show that the electric 16-pole and 64-pole 
moment operators \cite{Kusunose08}, 
expressed by the Racah operators \cite{Jensen91} 
$\tilde{O}_{p,q}({\bm J})$ with $p=4$ and 6, respectively 
(or Stevens's operators), 
are proportional to $\sigma^x \pm i \sigma^y$ or 
the unit matrix within $| \pm 1 \rangle_{\text{D}} $. 
Therefore within the CF ground state doublet, 
pseudospin operators $\sigma^x$ and $\sigma^y$ represent 
the electric multipole moments. 
A single-site CF ground state expressed by 
\begin{equation}
| \psi \rangle = (| 1 \rangle_{\text{D}}, | -1 \rangle_{\text{D}}) \chi, 
\label{GSs}
\end{equation}
where $\chi$ is the pseudospin wave-function, 
has the largest expectation of the magnetic dipole moment 
$|\langle \psi| \sigma^z | \psi \rangle| = 1$ 
(and $\langle \psi| \sigma^x | \psi \rangle = \langle \psi| \sigma^y | \psi \rangle = 0$) 
for $\chi = \begin{pmatrix} 1 \\ 0 \end{pmatrix}$ 
or $\chi = \begin{pmatrix} 0 \\ 1 \end{pmatrix}$. 
The other states expressed by 
\begin{equation}
\chi =  \begin{pmatrix} \cos \frac{\theta}{2} e^{-i \phi/2} \\ \sin \frac{\theta}{2} e^{i \phi / 2} \end{pmatrix}
\label{spinor}
\end{equation}
in which $\theta$ is in the range $0 < \theta < \pi $ 
have finite expectation values of the electric quadrupole moment operators; 
$\langle \psi| \sigma^x | \psi \rangle \ne 0$ 
and/or $\langle \psi| \sigma^y | \psi \rangle \ne 0$. 
These states have slightly deformed $f$-electron charge densities 
from that of the magnetic states with $\theta = 0$ or $\theta = \pi$. 
More specifically, the approximate $f$-electron 
charge density \cite{Kusunose08} of the state $| \psi \rangle$ 
is given by 
\begin{equation}
\langle \psi | \rho(\bm{r}) | \psi \rangle \simeq 
(-e) [R_f(r)]^2 \langle \psi |\rho_e(\hat{\bm{r}}) | \psi \rangle  \tfrac{1}{4 \pi}. 
\label{FCD1}
\end{equation}
The angular dependence \cite{Kusunose08} $\rho_e(\hat{\bm{r}})$ of this equation is
\begin{equation}
\rho_e(\hat{\bm{r}}) = n + \sum_{p=2,4,6; q } 
[4 \pi (2p+1)]^{1/2} \alpha_p Y_{p,q}(\hat{{\bm r}})^* \tilde{O}_{p,q}({\bm J}),
\label{FCD2}
\end{equation}
where $(\alpha_2, \alpha_4, \alpha_6) = (\alpha, \beta, \gamma)$ are 
the Stevens factors \cite{Jensen91}, 
$n=8$ is the number of $f$-electrons, and 
$Y_{p,q}(\hat{{\bm r}})$ are the spherical harmonics. 
By evaluating $\langle \psi |\rho_e(\hat{\bm{r}}) | \psi \rangle$ 
using several spinors of Eq.~(\ref{spinor}), 
one can show that the deformation of the $f$-electron charge density 
is mainly determined by the electric quadrupole moments. 
The electric 16-pole and 64-pole moments have non-negligible 
contributions to the deformation similarly to the analyses of 
the CF states \cite{Gingras00,Mirebeau07,Bertin12,Zhang14,Princep15}. 
In these meanings, 
the CF ground (psedospin-1/2) states $| \psi \rangle$ 
can be referred to as composite spin and quadrupole states. 

\section{Effective Pseudospin-1/2 Hamiltonian}
The generic form of the effective pseudospin-1/2 
Hamiltonian for non-Kramers CF ground state doublets of 
4$f$ magnetic ions 
in the pyrochlore structure was derived in Ref.~\cite{Onoda11} 
by calculating the nearest-neighbor (NN) superexchange interaction. 
This Hamiltonian consists of two parts. 
The first part is the NN magnetic interaction 
\begin{equation}
H_{\text{m,NN}} = J_{\text{nn}} 
\sum_{\langle {\bm r} , {\bm r}^{\prime} \rangle} 
\sigma_{\bm{r}}^{z} \sigma_{\bm{r}^{\prime}}^{z},
\label{HmagNN}
\end{equation}
which represents the NN classical spin-ice model for $J_{\text{nn}} > 0$.
The second part is the NN quadrupolar interaction 
\begin{eqnarray}
H_{\text{q}} = &J_{\text{nn}}& \sum_{\langle \bm{r} , \bm{r}^{\prime} \rangle} 
[ 2 \delta ( \sigma_{\bm{r}}^+ \sigma_{\bm{r}^{\prime}}^- 
           + \sigma_{\bm{r}}^- \sigma_{\bm{r}^{\prime}}^+ ) \nonumber \\
&+& 2q ( e^{2 i \phi_{\bm{r},\bm{r}^{\prime}} } 
\sigma_{\bm{r}}^+ \sigma_{\bm{r}^{\prime}}^+ + \text{H.c.} ) ],
\label{Hquad}
\end{eqnarray}
where $ \sigma_{\bm{r}}^{\pm} = ( \sigma_{\bm{r}}^x \pm i \sigma_{\bm{r}}^y ) / 2 $ 
and $ \sigma_{\bm{r}}^{\alpha}$ 
($\alpha = x, y, z $ defined using the local axes 
Eqs.~(\ref{Laxis0})-(\ref{Laxis3})) stand for 
the Pauli matrices of the pseudospin at a site $\bm{r}$.  
The phases $\phi_{\bm{r},\bm{r}^{\prime}}$ are 
$\phi_{\bm{r},\bm{r}^{\prime}} = 0$, $- 2 \pi /3$, and $2 \pi /3$ 
for $(i, i^{\prime}) = (0,3), (1,2)$, 
$(i, i^{\prime}) = (0,1), (2,3)$, 
and $(i, i^{\prime}) = (0,2), (1,3)$, respectively, 
where 
$( \bm{r} , \bm{r}^{\prime} ) = ( \bm{t}_n + \bm{d}_i , \bm{t}_{n^{\prime}} + \bm{d}_{i^{\prime}} ) $.

For the magnetic interaction of TTO we probably have to 
include the classical dipolar interaction, i.e., 
\begin{eqnarray}
&& H_{\text{m}} = H_{\text{m,NN}} + D r_{\text{nn}}^3 \nonumber \\
&\times& \sum_{\langle \bm{r} , \bm{r}^{\prime} \rangle} \left\{
\frac{ \bm{z}_{{\bm r}} \cdot \bm{z}_{{\bm r}^{\prime}} }{ | \Delta \bm{r} |^3} - 
 \frac{ 3 [\bm{z}_{{\bm r}} \cdot \Delta \bm{r} ] [\bm{z}_{{\bm r}^{\prime}} \cdot \Delta \bm{r} ] }{| \Delta \bm{r} |^5} \right\} \sigma_{\bm{r}}^{z} \sigma_{\bm{r}^{\prime}}^{z},
\label{Hmag}
\end{eqnarray}
where the summation runs over all pairs of sites, 
$r_{\text{nn}}$ is the NN distance, 
and $\Delta \bm{r} =  \bm{r} - \bm{r}^{\prime}$. 
The parameter $D$ is determined by 
the magnetic moment of the CF ground state doublet. 
We adopt $D = 0.29$ K, corresponding to 
the experimental value of the magnetic 
moment 4.6 $\mu_{\text{B}}$ \cite{Takatsu15}. 
As discussed in Refs.~\cite{Hertog00,Isakov05}, 
when the magnetic interaction of Eq.~(\ref{Hmag}) represents the dipolar 
spin ice ($J_{\text{nn}} + D_{\text{nn}} > 0$), 
$H_{\text{m}}$ can be approximated by the 
NN classical spin-ice Hamiltonian \cite{Hertog00} 
\begin{eqnarray}
H_{\text{m}} \simeq ( J_{\text{nn}} + D_{\text{nn}} )
\sum_{\langle \bm{r} , \bm{r}^{\prime} \rangle} 
\sigma_{\bm{r}}^{z} \sigma_{\bm{r}^{\prime}}^{z},
\label{HmagSI}
\end{eqnarray}
where $D_{\text{nn}} = \frac{5}{3} D = 0.48$ K.

In our computations we used 
an effective pseudospin-1/2 Hamiltonian of  
the form 
\begin{equation}
H_{\text{eff}} = H_{\text{m}} +  H_{\text{q}}.
\label{Heff}
\end{equation}
We note that this is not very different from 
the original Onoda-type interaction \cite{Onoda11} 
($D_{\text{nn}}=0$) and results of 
Refs.~\cite{Onoda11,Lee12} 
can be approximately used at least 
in the electric quadrupolar phases, 
in which $xy$-components of the pseudospin 
$(\sigma_{\bm{r}}^x, \sigma_{\bm{r}}^y)$ show LRO and 
semi-classical theoretical treatments are applicable. 

\section{Pseudopin Wave}
The studies \cite{Onoda11,Lee12,Takatsu15} of the effective 
Hamiltonian of Eq.~(\ref{Heff}) showed 
that there are two electric quadrupolar states: 
the PAF state (planar antiferropseudospin) 
and 
the PF state (planar ferropseudospin) 
depending on the two parameters $(\delta , q )$ 
(see Fig. 7 in Ref.~\cite{Onoda11} and 
Fig. 3 in Ref.~\cite{Lee12} for details). 
In these states, the $xy$-components of the pseudospin 
show LRO with the modulation vector $\bm{k} = 0$. 
It should be noted that this wave vector $\bm{k} = 0$ is selected 
by quantum \cite{Lee12} and thermal \cite{Onoda11,Takatsu15} 
fluctuations for PAF, i.e., by an order-by-disorder mechanism. 

In order to calculate elementary excitations 
in the PAF and PF states, 
we choose one of the pseudospin structures 
\begin{equation}
(\langle \sigma_{ \bm{t}_n + \bm{d}_i }^{x} \rangle,\langle \sigma_{ \bm{t}_n + \bm{d}_i }^{y} \rangle) = 
\begin{cases} 
(0,\langle \sigma^{y} \rangle) & (i=0,3) \\
-(0,\langle \sigma^{y} \rangle) & (i=1,2) \; (\text{PAF}) 
\end{cases}
\label{PAF}
\end{equation}
and 
\begin{equation}
(\langle \sigma_{ \bm{t}_n + \bm{d}_i }^{x} \rangle,\langle \sigma_{ \bm{t}_n + \bm{d}_i }^{y} \rangle) = 
(0,\langle \sigma^{y} \rangle) \; (i=0,1,2,3) \; (\text{PF}). 
\label{PF}
\end{equation}
We apply the simple linear spin-wave theory, MF-RPA \cite{Jensen91} 
(mean field, random phase approximation), 
in the same way as described in \S 3.5.2 of Ref.~\cite{Jensen91}. 
In MF-RPA, $\langle \sigma^{y} \rangle$ of Eqs.~(\ref{PAF}) and (\ref{PF}) 
is calculated by the MF approximation. 
For the present purpose, 
we are interested in elementary excitations only at low temperatures, 
and $\langle \sigma^{y} \rangle = 1$ is a good approximation. 
To obtain dispersion relations of pseudospin waves, 
MF-RPA utilizes the generalized susceptibility $\chi(\bm{k}, E)$ 
and neutron magnetic scattering intensity $S(\bm{Q},E)$ \cite{Jensen91}. 
Useful examples of MF-RPA computations 
including straightforward technical extensions for pyrochlore structures 
are described Refs.~\cite{Kao03,Petit12TSO}. 
General computational treatments of MF-RPA are discussed 
in Refs.~\cite{Rotter06,McPhase2014}. 
Following these references \cite{Jensen91,Kao03,Petit12TSO,McPhase2014}, 
the generalized susceptibility is given by 
\begin{equation}
\chi(\bm{k}, E) = [1 - \chi^{0}(E) J(\bm{k})]^{-1} \chi^{0}(E).
\label{Gsus}
\end{equation}
where $\bm{k}$ is a wave vector in the first Brillouin zone, 
$\chi^{0}(E)$ and $J(\bm{k})$ denote the single-site 
generalized-susceptibility of the MF Hamiltonian and 
the Fourier transform of the exchange and dipolar coupling constants. 
The neutron magnetic scattering intensity $S(\bm{Q}=\bm{G}+\bm{k},E)$ 
is given by 
\begin{eqnarray}
&S&(\bm{Q},E) \propto \frac{1}{1 - e^{-\beta E}} 
\sum_{\rho,\sigma} (\delta_{\rho,\sigma} - \hat{Q}_{\rho} \hat{Q}_{\sigma} ) \nonumber \\
&\times&\sum_{i,i^{\prime}} 
U_{\rho,z}^{(i)} U_{\sigma,z}^{(i^{\prime})} 
\text{Im} \left\{ 
\chi_{i,z;i^{\prime},z}(\bm{k}, E) 
e^{- i \bm{G} \cdot (\bm{d}_{i} - \bm{d}_{i^{\prime}}) } 
\right\},
\label{SQE}
\end{eqnarray}
where only the local $z$-component of the pseudospin $\sigma_{\bm{r}}^z$ 
contribute to the scattering. 
If one assumes that all the pseudospin components represent 
a magnetic dipole moment vector with an isotropic $g$-factor, 
virtual neutron scattering intensity $S_{\text{v}}(\bm{Q},E)$ is given by 
\begin{eqnarray}
&S_{\text{v}}&(\bm{Q},E) \propto \frac{1}{1 - e^{-\beta E}} 
\sum_{\rho,\sigma} (\delta_{\rho,\sigma} - \hat{Q}_{\rho} \hat{Q}_{\sigma} ) \nonumber \\
&\times&\sum_{i,\alpha,i^{\prime},\alpha^{\prime}} 
U_{\rho,\alpha}^{(i)} U_{\sigma,\alpha^{\prime}}^{(i^{\prime})} 
\text{Im} \left\{ 
\chi_{i,\alpha;i^{\prime},\alpha^{\prime}}(\bm{k}, E) 
e^{- i \bm{G} \cdot (\bm{d}_{i} - \bm{d}_{i^{\prime}}) } 
\right\},
\label{SQEall}
\end{eqnarray}
where $U_{\rho,\alpha}^{(i)}$ is the rotation matrix \cite{Kao03,Petit12TSO} 
from the local ($\alpha$) frame defined at the sites 
$\bm{t}_n + \bm{d}_i$ to the global ($\rho$) frame. 
$S_{\text{v}}(\bm{Q},E)$ is useful when displaying 
dispersion relations of all pseudospin waves, 
because the amplitude of the electric quadrupole 
moment are excluded for $S(\bm{Q},E)$. 

\begin{figure}
\centering
\includegraphics[width=8.0cm]{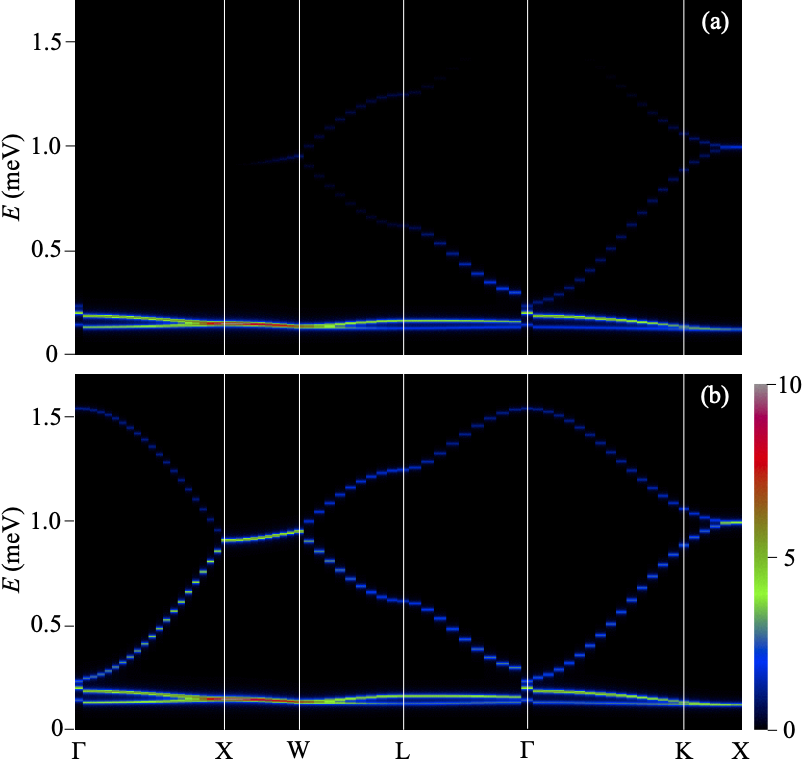}
\caption{
Magnetic $S(\bm{Q},E)$ (a) and 
virtual $S_{\text{v}}(\bm{Q},E)$ (b) 
of the PAF ordering (Eq.~(\ref{PAF})) using interaction parameters 
$J_{\text{nn}}=1$ K, $q=0.85$, and $\delta=0$.}
\label{fig1}
\end{figure}

\begin{figure}
\centering
\includegraphics[width=8.0cm]{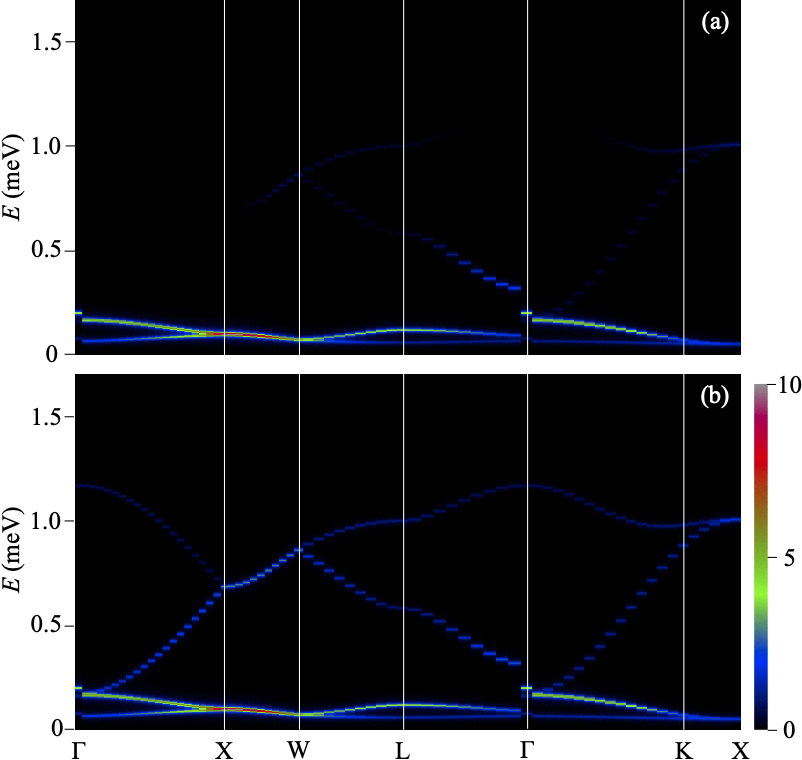}
\caption{
Magnetic $S(\bm{Q},E)$ (a) and 
virtual $S_{\text{v}}(\bm{Q},E)$ (b) 
of the PAF ordering (Eq.~(\ref{PAF})) using interaction parameters 
$J_{\text{nn}}=1$ K, $q=0.5$, and $\delta=0.6$.}
\label{fig2}
\end{figure}

\begin{figure}
\centering
\includegraphics[width=8.0cm]{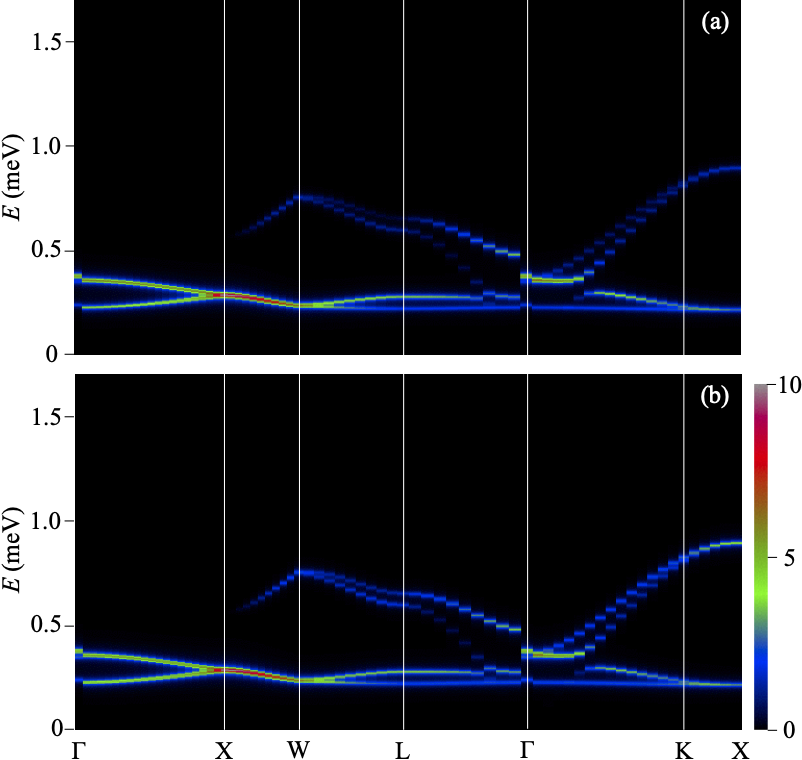}
\caption{
Magnetic $S(\bm{Q},E)$ (a) and 
virtual $S_{\text{v}}(\bm{Q},E)$ (b) 
of the PF ordering (Eq.~(\ref{PF})) using interaction parameters 
$J_{\text{nn}}=1$ K, $q=0.8$, and $\delta=-0.6$.}
\label{fig3}
\end{figure}

\begin{figure}
\centering
\includegraphics[width=8.0cm]{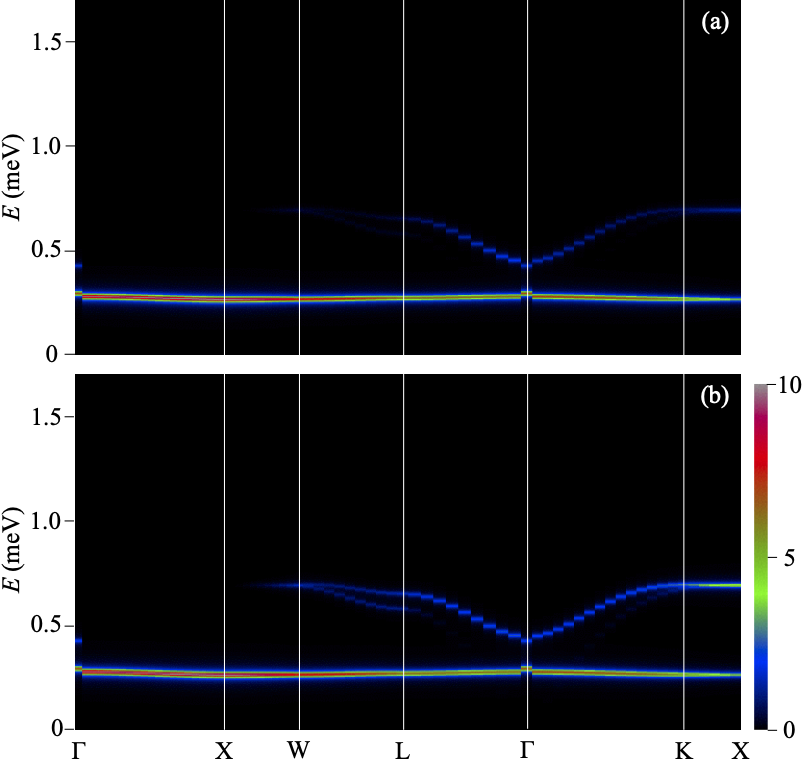}
\caption{
Magnetic $S(\bm{Q},E)$ (a) and 
virtual $S_{\text{v}}(\bm{Q},E)$ (b) 
of the PF ordering (Eq.~(\ref{PF})) using interaction parameters 
$J_{\text{nn}}=1$ K, $q=0$, and $\delta=-0.6$.}
\label{fig4}
\end{figure}

In Fig.~\ref{fig1}(a) we show the inelastic magnetic 
scattering intensity $S(\bm{Q},E)$ (Eq.~(\ref{SQE})) 
of the the PAF ordering 
(Eq.~(\ref{PAF})) along several symmetry directions 
in the FCC Brillouin zone 
using the interaction parameters $J_{\text{nn}}=1$ K, $q=0.85$, 
and $\delta=0$ adopted in Ref.~\cite{Takatsu15}. 
One can see two flat excitation branches in Fig.~\ref{fig1}(a). 
We also show the virtual $S_{\text{v}}(\bm{Q},E)$ (Eq.~(\ref{SQEall})) 
in Fig.~\ref{fig1}(b). 
This figure clearly shows that there are four excitation 
branches consistent with the $\bm{k}=0$ structure 
posessing four sites in the unit cell. 
These four pseudospin-wave branches are composite 
spin ($\sigma_{\bm{r}}^z$) and quadrupole ($\sigma_{\bm{r}}^x$) 
waves. 
Figs.~\ref{fig1}(a) and (b) show that the amplitude of 
the spin components is strong and weak 
in the two lower-$E$ and the two higher-$E$ branches, respectively. 
Fig.~\ref{fig2} shows the magnetic $S(\bm{Q},E)$ 
and virtual $S_{\text{v}}(\bm{Q},E)$ 
using different parameters 
$J_{\text{nn}}=1$ K, $q=0.5$, and $\delta=0.6$ 
in the PAF phase (Eq.~(\ref{PAF})). 
The two lower-$E$ excitation branches become more dispersive 
by the finite value of $\delta$ compared to Fig.~\ref{fig1}. 

In Fig.~\ref{fig3} we show the magnetic $S(\bm{Q},E)$ 
and virtual $S_{\text{v}}(\bm{Q},E)$ 
using parameters 
$J_{\text{nn}}=1$ K, $q=0.8$, and $\delta=-0.6$, 
which are in the PF phase (Eq.~(\ref{PF})). 
Compared to the PAF cases, the difference between 
the magnetic and virtual $S(\bm{Q},E)$ 
becomes less pronounced. 
Fig.~\ref{fig4} shows the magnetic $S(\bm{Q},E)$ 
and virtual $S_{\text{v}}(\bm{Q},E)$ 
using parameters 
$J_{\text{nn}}=1$ K, $q=0$, and $\delta=-0.6$ 
in the PF phase (Eq.~(\ref{PF})). 
For vanishing $q=0$, the two lower-$E$ branches are more 
flattened and merge into almost one branch.

\section{Magnetic Spectra of Polycrystalline Tb$_{2+x}$Ti$_{2-x}$O$_{7+y}$}
\begin{figure}
\centering
%
\includegraphics[width=8.0cm]{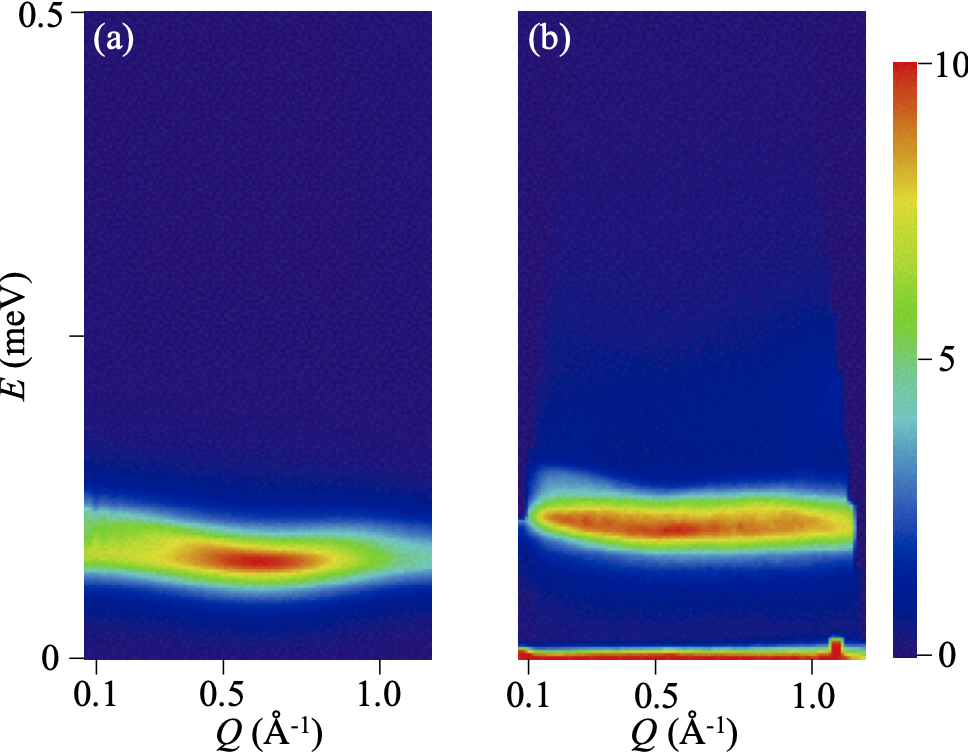}
\caption{
(a) Powder averaged magnetic $S(|\bm{Q}|,E)$ 
of the PAF ordering (Eq.~(\ref{PAF})) 
using interaction parameters 
$J_{\text{nn}}=1$ K, $q=0.8$, and $\delta=0$. 
(b) Inelastic neutron scattering spectra of 
polycrystalline Tb$_{2+x}$Ti$_{2-x}$O$_{7+y}$ 
with $x = 0.005$ at $T=0.1$ K well below $T_{\text{c}}$.}
\label{fig5}
\end{figure}

Finally, we would like to compare the previously 
observed \cite{Taniguchi13} inelastic magnetic neutron scattering spectra 
peaked around $E = 0.1$ meV 
with the present pseudospin wave calculation. 
The sample is the polycrystalline Tb$_{2+x}$Ti$_{2-x}$O$_{7+y}$ 
with $x = 0.005$ ($T_{\text{c}} = 0.5$ K) \cite{Taniguchi13}. 
The neutron scattering experiment was performed 
on the time-of-flight spectrometer ILL-IN5 operated 
with $\lambda = 10$ {\AA}. 
Fig.~\ref{fig5}(b) shows $Q$-dependent powder spectra taken 
at $T = 0.1$ K. 
These data should be compared with powder averaging of 
the magnetic $S(\bm{Q},E)$. 
Fig.~\ref{fig5}(a) shows an example of this powder 
averaged $S(|\bm{Q}|,E)$ choosing the parameters 
$J_{\text{nn}}=1$ K, $q=0.8$, and $\delta=0$, 
which are in the PAF phase (Eq.~(\ref{PAF})). 
We think that these figures show reasonably good agreement 
between the calculation and the observation. 
In spite of using the over-simplified model Hamiltonian 
for TTO and the crude linear-spin-wave theory for 
the frustrated quantum system, 
essential features of experimental spectra can be reproduced 
by the approximate calculation. 
The slight $Q$-dependence and the non-resolution limited 
peak-width, $\Delta E \gg (\Delta E)_{\text{resolution}}=0.01 $ meV, 
have been one of the puzzling observations of TTO. 
The present interpretation using the composite spin-quadrupole 
wave can be an answer \cite{Takatsu15,Wakita15}.

\section{Summary}
In this study, we try to reformulate the problem of 
Tb$_{2+x}$Ti$_{2-x}$O$_{7+y}$ and reinterpret its 
puzzling experimental facts 
based on the theoretically predicted \cite{Onoda11} 
pseudospin-1/2 Hamiltonian including the electronic 
superexchange interaction between electric quadrupole moments. 
In this scenario, the hidden order 
in some TTO samples is an electric quadrupolar LRO. 
Although this LRO does not give rise to 
strong magnetic Bragg scattering, it can be 
observed by inelastic magnetic neutron scattering 
as a composite spin-quadrupole wave. 
We employ a MF-RPA linear spin-wave theory and 
compare its computation with previously observed 
low-energy magnetic excitation spectra of 
a polycrystalline sample with $x = 0.005$ 
($T_{\text{c}} = 0.5 $ K). 
Quite intriguingly, the interaction parameters 
used in Fig.~\ref{fig5}(a) are located very close to 
the phase boundary between the PAF and 
U(1) quantum spin-liquid states \cite{Lee12,Onoda11,Takatsu15}. 
This may possibly imply that Tb$_{2+x}$Ti$_{2-x}$O$_{7+y}$ 
samples with $x < x_{\text{c}}$ are 
in the U(1) quantum spin-liquid phase \cite{Takatsu15}.
\begin{acknowledgments}
We thank S. Onoda and Y. Kato 
for useful discussions. 
This work was supported by JSPS KAKENHI grant numbers 25400345 and 26400336. 
The neutron scattering performed using ILL-IN5 (France) 
was transferred from JRR3-HER (proposal 11567) 
with the approval of ISSP, Univ. of Tokyo, and JAEA, Tokai, Japan.
\end{acknowledgments}

\bibliography{TTO_spin_kado}
\end{document}